\documentclass[a4paper]{article}

\usepackage{INTERSPEECH2019}

\title{Self-Teaching Networks}
\name{Liang Lu, Eric Sun and Yifan Gong}
\address{Microsoft}
\email{\{liang.lu, sun.eric, yifan.gong\}@microsoft.com}

\begin{document}

\maketitle
\begin{abstract}
We propose self-teaching networks to improve the generalization capacity of deep neural networks. The idea is to generate soft supervision labels using the output layer for training the lower layers of the network. During the network training, we seek an auxiliary loss that drives the lower layer to mimic the behavior of the output layer. The connection between the two network layers through the auxiliary loss can help the gradient flow, which works similar to the residual networks. Furthermore, the auxiliary loss also works as a regularizer, which improves the generalization capacity of the network. We evaluated the self-teaching network with deep recurrent neural networks on speech recognition tasks, where we trained the acoustic model using 30 thousand hours of data. We tested the acoustic model using data collected from 4 scenarios. We show that the self-teaching network can achieve consistent improvements and outperform existing methods such as label smoothing and confidence penalization. 
\end{abstract}

\section{Introduction}

Deep neural networks trained with millions of parameters have achieved tremendous success in various machine learning tasks including speech, natural language and image processing~\cite{LeCun2015}. Despite the success, there are still some challenges to be address in deep learning. For example, although trained with huge amount of data, the neural networks still cannot generalize well to the scenarios which are not well represented by the training data, and overfit to domains which dominate the training set. One example is speech recognition in the `wild', where a single speech recognizer is deployed in various acoustic conditions and domains including far- or near-field and different dialects. Due to the vanishing gradient problem, another challenge is to train deeper networks that can fully harness the information from the vast amount of data. Residual and highway networks~\cite{he2015deep, srivastava2015training} have made significant progress along this direction, however, their applications in recurrent networks are not as successful as in feedforward and convolutional networks. Training very deep networks with high generalization capacity is still a challenging optimization problem in the research community.

There have been significant research efforts devoted to improve the generalization ability of the deep neural networks. Traditional approaches include the use of weight noise, L1/L2 regularization, and dropout~\cite{srivastava2014dropout}. However, these approaches are not widely used in training commercial speech recognition systems where the amount of training data is in thousands to tens of thousands hours, as their effects tend to diminish when the amount of data increases. Recently,  a few regularization approaches for neural networks have been proposed that introduce noise to the supervision labels, e.g., label smoothing~\cite{Szegedy_2016_CVPR} and confidence penalization~\cite{Pereyra:2017aa}. The rationale behind these approaches is that in the supervised learning task, the ground truth labels are from a Multinoulli distribution with the probability of the correct class being 1 and others being 0.  However, for neural networks which use softmax as the output layer, the predictions are soft probabilities, which means they cannot completely match the ground truth labels. Training the network with more epochs can always push down the cross-entropy loss, but with the cost of overfitting to the training set, indicating a lower generalization capacity. To preventing overfitting and improve generalization, a random noise is applied to the labels in label smoothing, allowing the model to match an `easier' target distribution. Motivated in a similar way, confidence penalization aims at penalizing the entropy of the output distribution of the neural network to avoid the network placing the entire probability to a single class. 

In this paper, we propose self-teaching networks -- another approach to tackle the vanishing gradient problem in deep networks as well as improving the generalization capacity of the network. In contrast to residual and highway networks, which feed the features from the lower layers into the upper layers with skip connections, in self-teaching networks, we pass the supervision signal from the top layer down to the lower layer directly. To this end, we use the top layer to generate the supervision labels, and use them as the additional supervision signal to train the lower layers. This is analogous to teacher-student training~\cite{bucilu2006model, li2014learning, hinton2015distilling}, where the teacher model generates the soft labels to train the student model. But the key difference here is that the `teacher' and `student' now refer to different layers of the {\it same} network. Thus we name this approach as `self-teaching'. Given the soft supervision labels from the output layer, we seek an auxiliary loss to minimize the error compared to the soft labels. While it benefits the gradient flow due to the skip connection, the auxiliary loss also works as a regularizer, which improves the generalization capacity of the network. While this is a general training approach applicable to different types of network architectures, in this study, we narrow down the scope and focus on deep recurrent networks with long short-term memory (LSTM) units~\cite{hochreiter1997long}. In terms of applications, we evaluate the self-teaching networks on speech recognition tasks, where we trained the acoustic model using 30 thousand hours of data. Our test dataset consists of audio recorded from 4 different scenarios, and we show consistent improvements by the self-teaching network as well as better generalization compared to label smoothing and confidence penalization.    


\section{Method}
\label{sec:self}

When the neural network goes deeper, the gradients to the lower layers become noisy and less informative. An indication of this behavior is the underfitting phenomena as observed in~\cite{he2015deep}. In fact, Zhang et al.~\cite{Zhang2019} shows that the lower layers of the network may be more important to extract the useful feature representations in classification tasks. Residual and highway networks~\cite{he2015deep, srivastava2015training} tackle this problem by introducing skip connections which feed the lower level features directly to the upper layers. In this case, the lower layers will be closer to the supervision signals and they can receive higher quality gradient signals for parameter update. 

\begin{figure}[t]
\small
\centerline{\includegraphics[width=0.25\textwidth]{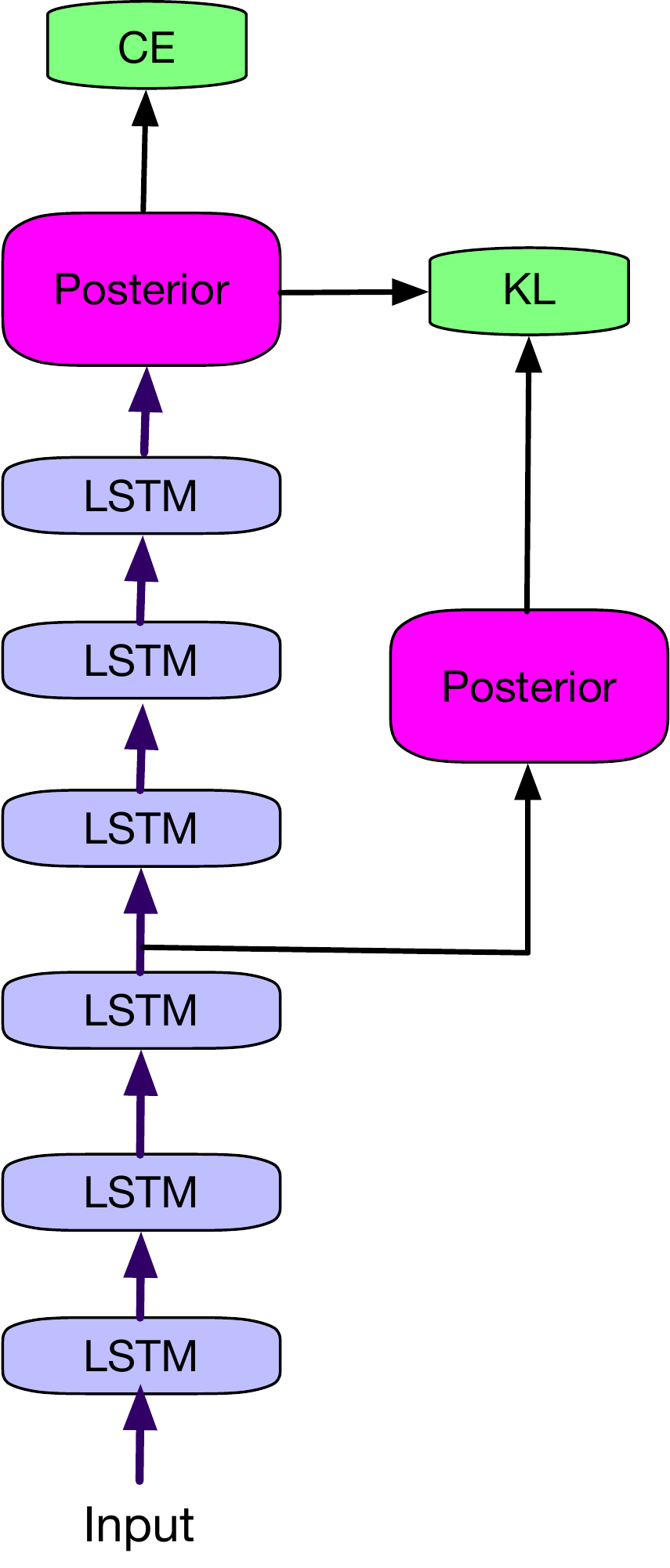}} \vskip -2mm
\caption{An example of self-teaching network with a 6-layer LSTM studied in this work. If the posterior is at the frame-level, we use cross-entropy (CE) as the main loss function. The auxiliary connection is from the top layer to the 3rd layer of the network.}  
\label{fig:self-teach}
\vskip -2mm
\end{figure}

Motivated in the similar way, we also draw connections between the top and lower layers in self-teaching networks, but it works the other way round. Rather than directly feeding the lower layer features upward through a skip connection, we pass down the supervision signal from the output layer to the lower layers. More precisely, suppose the posterior probability from the output layer of the network is $P_{\theta}(\bm{y}|\bm{h}^L)$, where $\bm{y}$ denotes the output, $\bm{h}^L$ is the activation from the top layer of the network, $L$ denotes the network depth and $\theta$ is the set of model parameters. We then add another softmax layer to a lower layer of the network to compute the posterior probability $P_{\theta}({\bm{y}|\bm{h}^l})$, where $1\le l < L$. During the network training, we also minimize the distance between the two distributions together with the original cross-entropy loss function in order to encourage the lower layer to follow the behaviour of the top layer. If we use the Kullback-Leibler (KL) divergence as the measure between the two posterior distributions, the loss function becomes
\begin{align}
\mathcal{L}(\theta) &= H\left(\bm{\bar{y}}, P_{\theta}(\bm{y}|\bm{h}^L)\right) + \lambda D_{KL}\left(P_{\theta}(\bm{y}|\bm{h}^L)||P_{\theta}(\bm{y}|\bm{h}^l)\right) \nonumber \\
&= H\left(\bm{\bar{y}}, P_{\theta}(\bm{y}|\bm{h}^L)\right) + \lambda H\left(P_{\theta}(\bm{y}|\bm{h}^L), P_{\theta}(\bm{y}|\bm{h}^l)\right)  \nonumber \\
\label{eq:self-teach}
& - \lambda H(P_{\theta}(\bm{y}|\bm{h}^L) )
\end{align}
where $\bm{\bar{y}} \in \{0, 1\}^Z, \sum_j \bar{\bm y}_j = 1$ is the ground truth label with the number of classes as $Z$; $\lambda$ is the tunable parameter; $H(\bm{p})$ is the entropy of the distribution $\bm{p}$,
\begin{align}
H(\bm{p}) = - \sum_i \bm{p}_i \log \bm{p}_i,  
\end{align}
and  $H(\bm{p}, \bm{q})$ denotes the cross-entropy of the two distributions $\bm{p}$ and $\bm{q}$, i.e.,
\begin{align}
H(\bm{p}, \bm{q}) = - \sum_i \bm{p}_i \log \bm{q}_i 
\end{align}
for discrete distributions. $D_{KL}(\bm{p}||\bm{q})$ denotes the KL-divergence between the two distributions, i.e.,
\begin{align}
D_{KL}(\bm{p}, \bm{q}) = H(\bm{p}, \bm{q}) - H(\bm{p}).
\end{align} 
An example of the self-teaching network with LSTMs is shown in Figure~\ref{fig:self-teach}, which we will focus on in this paper. 

To compute the probability $P_{\theta}({\bm{y}|\bm{h}^l})$, it requires an additional linear projection layer to compute the logits (unnormalized probabilities) before the softmax operation. If the number of classes $Z$ is large, this can results in considerable increase in terms of the model size during training, although this layer is discarded during inference, which means there is no additional computational cost to deploy the model. When the amount of training data is relatively small, it may be reasonable to share the linear projection layer with the output layer to reduce the total number of model parameters in training. However, this is not investigated as we have abundant training data as detailed in the experimental section. 

For speech recognition, the posterior probabilities $P_{\theta}({\bm{y}|\bm{h}^L})$ and $P_{\theta}({\bm{y}|\bm{h}^l})$ can be at the frame-level or at the sequence-level. In the frame-level scenario, the posterior probabilities are obtained simply by feeding in the acoustic frames to the network followed by the softmax operation. The loss function is in the form of cross-entropy. The loss function corresponding to the self-teaching network in this case is shown by Eq~\eqref{eq:self-teach}. For the case of sequence-level scenario, the two probabilities can be obtained by forward-backward computation over the lattices, and the main cross-entropy loss function $H\left(\bm{\bar{y}}, P_{\theta}(\bm{y}|\bm{h}^L)\right)$ should be replaced by the one corresponding to maximum mutual information (MMI) or minimum Bayesian risk (MBR), while the KL-divergence term remains the same. In this work, we only focus on the frame-level scenario, and leave the sequence-level training as our future work. 

\section{Related Work}
\label{sec:related}

As mentioned above, the self-teaching network proposed in this paper is related to several previous works in the literature. In this section, we only briefly review a few closely related ones.

\subsection{Label smoothing}

Label smoothing~\cite{Szegedy_2016_CVPR} is a simple method to add noise to the supervision labels to increase the robustness of the network. It can be viewed as sampling from a noise distribution $\bm q$ and adding the output to the ground truth labels $\bm{\bar{y}}$. The noise perturbed labels are then renormalized to follow a valid distribution, which are then used as the new supervision labels. As shown in~\cite{Szegedy_2016_CVPR}, it is equivalent to minimize the follow loss function:
\begin{align}
\mathcal{L}(\theta) &=  H\left(\bm{\bar{y}}, P_{\theta}(\bm{y}|\bm{h}^L)\right)  + \lambda D_{KL}\left(\bm{q}||P_{\theta}(\bm{y}|\bm{h}^L)\right). 
\end{align}
The noise distribution $\bm{q}$ can be uniform as used in~\cite{Szegedy_2016_CVPR, chiu2018, Kim:2017aa}, or a Multinoulli distribution as in~\cite{Xie_2016_CVPR}. Compared to the self-teaching network as Eq~\eqref{eq:self-teach}, the key difference is that the noise distribution is replaced by the posterior computed from a lower layer output of the network. 

\subsection{Confidence penalization}

Confidence penalization~\cite{Pereyra:2017aa} aims at penalizing the output distribution of the network which places entire probability to a single class. The method is to introduce a penalization term that encourages the output distribution to have a higher entropy. From the information theory, a distribution reaches its maximum entropy when it is uniform. Consequently, the effect of the penalization term is to push the model to place reasonable probabilities to the entire classes, so that it will be closer to a uniform distribution. This is in effect similar to label smoothing with the uniform label noise. Concretely, the loss function of confidence penalization can be expressed as
\begin{align}
\mathcal{L}(\theta) &=  H\left(\bm{\bar{y}}, P_{\theta}(\bm{y}|\bm{h}^L)\right)  - \lambda H\left(P_{\theta}(\bm{y}|\bm{h}^L)\right).
\end{align}
Compared to Eq~\eqref{eq:self-teach}, it is the same as the loss function of the self-teaching network, but without the cross-entropy term between $P_{\theta}(\bm{y}|\bm{h}^L)$ and $P_{\theta}(\bm{y}|\bm{h}^l)$.
 
\subsection{Teacher-Student learning}

The self-teaching network is also related to the teacher-student learning~\cite{li2014learning, hinton2015distilling} approach for transfer learning between different networks. Let $\bar{\theta}$ be the set of parameters of the teacher model, the loss function for teacher-student learning may be represented as
\begin{align}
\label{eq:TS}
\mathcal{L}(\theta) &= H\left(\bm{\bar{y}}, P_{\theta}(\bm{y}|\bm{h}^L)\right) + \lambda D_{KL}\left(P_{\bar{\theta}}(\bm{y}|\bm{h}^L)||P_{\theta}(\bm{y}|\bm{h}^L)\right).
\end{align}
Since the teacher model $\bar{\theta}$ is not updated during the teacher-student training, the entropy term $H(P_{\bar{\theta}}(\bm{y}|\bm{h}^L))$ is ignored during the gradient computation. Hence the loss can be simplified as 
\begin{align}
\mathcal{L}(\theta) &= H\left(\bm{\bar{y}}, P_{\theta}(\bm{y}|\bm{h}^L)\right) + \lambda H\left(P_{\bar{\theta}}(\bm{y}|\bm{h}^L), P_{\theta}(\bm{y}|\bm{h}^L)\right).
\end{align}
In practice, the cross entropy from the ground truth labels $H\left(\bm{\bar{y}}, P_{\theta}(\bm{y}|\bm{h}^L)\right)$ is also ignored as the it does not make much difference in both convergence and accuracy~\cite{lu2016knowledge}. Without this term, it also allows the unlabeled data to be used in teacher-student training, which is a particular advantage. Comparing Eq~\eqref{eq:TS} to Eq~\eqref{eq:self-teach}, the formulation of the self-teaching network is the same as that of the teacher-student learning, while the major difference is that both the teacher and student model are different components or layers from the same model in the self-teaching network.

\subsection{Residual and highway networks}

The self-teaching network shares some similarities with residual and highway networks~\cite{he2015deep, srivastava2015training} in the sense that they all aim at tackling the vanishing gradient problem when propagating back the gradients to the lower layers of the network. They all create skip connections so that the lower layers are closer to the supervision signals. However, for self-teaching networks, the skip connection is to pass down the supervision signal from the output layer, while for residual and highway networks, the skip connection is used to feed the lower layer features directly to the upper layers. 

\section{Experiments and Results}
\label{sec:exp}

\subsection{Experimental setup}

We evaluated the self-teaching networks for speech recognition tasks. We focus on LSTMs for acoustic modeling as they are still the backbone for most of the state-of-the-art systems. In our experiments, the models were trained with 30 thousand hours of anonymized and transcribed Microsoft production data, recorded in both close-talk and far-field conditions. We work on hybrid systems, where the hidden Markov models (HMM) is used for sequence modeling. The number of tied triphone states is 9404 in our experiments, and the input feature is 80-dimension log Mel filter banks sampled every 20 milliseconds. The language model is a 5-gram with around 100 million (M) n-grams.  

Given the vast amount of training data, it is very expensive to thoroughly evaluate the two key aspects of the self-teaching network, namely, its abilities to improve the generalization capacity and to help gradient flow for very deep networks. In this study, we mainly focus on the generalization aspect of the self-teaching network. Therefore, we used 6-layer LSTMs for all our experiments as shown in Figure~\ref{fig:self-teach}, and only compare to the other two regularization approaches, i.e., label smoothing and confidence penalization. All the LSTM models were trained with truncated back-propagation through time (BPTT) with truncation length of 16. The acoustic models were evaluated on 4 test sets, which are detailed in Table~\ref{tab:dataset}. In particular, the {\tt DMA-O} condition is not covered by the training data, which is primarily for evaluating the generalization power of the model.

\begin{table}[t]
  \caption{Statistics of the 4 evaluation sets used in our experiments. {\tt Cortana} contains both close-talk and far-field speech. {\tt DMA-I} refers to the in-domain distant microphone array condition, and {\tt DMA-O} is the out-domain distant microphone array condition, which is not covered in the training data. }
  \label{tab:dataset}
  \centering
  \begin{tabular}{l|cc}
    \toprule
    Domain      & \# utterances          & \# words       \\
    \hline
    {\tt Cortana}                    & 79105        & 438970                               \\
    {\tt DMA-O}                  & 5485                 & 29480                \\
    {\tt DMA-I}                    & 75056                & 284092               \\
    \bottomrule
  \end{tabular}
\end{table}

\subsection{Results of smaller models}

Table~\ref{tab:small} shows the results of a smaller 6-layer LSTM model in terms of word error rates (WERs), where the size of the cell is 1024 and the dimension of the hidden vector is 512 after a linear projection layer. The number of the total model parameters is roughly 31 million. We trained the model using the CNTK deep learning toolkit~\cite{yu2014introduction}, and it took around 7 -- 10 days with parallel training using 16 GPUs for each model to converge. We searched the hyper-parameter $\lambda \in \{0.001, 0.005, 0.01, 0.02\}$, and for clarity, we only show results with the optimal $\lambda$. In fact, we observed a consistent trend by tuning the hyper-parameter $\lambda$. While label smoothing has been shown to improve the accuracy for end-to-end models such as the encoder-decoder attention model in~\cite{chiu2018}, and Connectionist Temporal Classification (CTC) model in~\cite{Kim:2017aa}, we did not observe consistent improvements for our hybrid model by varying the value of $\lambda$. Confidence penalization shows the similar trend. For self-teaching networks, however, we observed small but consistent improvements across the 4 evaluation sets. The new model is also fairly robust to the hyper-parameter in our experiments, as the results only fluctuated slightly when $\lambda$ is within the range from 0.001 to 0.02. Since the difference between confidence penalty and self-teaching network is only the cross entropy term $H\left(P_{\theta}(\bm{y}|\bm{h}^L), P_{\theta}(\bm{y}|\bm{h}^l)\right)$, we evaluated its impact by removing the entropy term $H(P_{\theta}(\bm{y}|\bm{h}^L) )$ in the experiments of self-teaching networks. As shown in Table~\ref{tab:small}, we can still observe consistent improvements without the entropy term, indicating that the cross-entropy term may be more important. This may help explain the results of confidence penalization, which show that the entropy term alone cannot achieve consistent WER reductions.  

\begin{table}[t]
  \caption{Experimental results of the smaller LSTM model with around 31 million parameters. {\tt LS} refers to label smoothing, and {\tt CP} is short for confidence penalization. {\tt ST} refers to the self-teaching network. {\tt H} means the entropy term $H(P_{\theta}(\bm{y}|\bm{h}^L) )$ in Eq~\eqref{eq:self-teach}. }
  \label{tab:small}
  \centering
  \begin{tabular}{l|ccc}
    \toprule
          & \multicolumn{3}{c}{WER}     \\
     System & {\tt Cortana} & {\tt DMA-O} & {\tt DMA-I} \\
    \hline
    Baseline                 & 9.46        & 19.56 & 13.37                               \\
    {\tt LS}     & 9.56        & 19.35  & 13.27               \\
    {\tt CP}             & 9.40 & 19.44  & 13.53                  \\
    {\tt ST} with {\tt H}      & 9.32 & 19.13 & 13.29                      \\
    {\tt ST} without {\tt H} & 9.35 & 19.37  & 13.18\\
    \bottomrule
  \end{tabular}
\end{table}

\subsection{Results of larger models}

We then increased the size of the model with the size of cell being 1800 and the dimension of the hidden vector being 600. The total number of model parameters becomes roughly 63 million, which is about twice as many as the model has in the previous experiments. This model takes around 2 weeks to converge when trained with 16 GPUs in parallel. Due to the computational cost, we only run experiments with self-teaching networks, and the results are shown in Table~\ref{tab:large}. The large model improved the baseline system, with WER reductions vary from 3.8\% for the {\tt Cortana} task to 4\% - 8\% for the {\tt DMA} task. For self-teaching networks, compared to the results of the smaller model, we observed relatively larger improvements after increasing the model size with the relative WER reduction up to 3\% -- 4\%, which indicates that the regularization term plays a larger role in the model training. Notably, the improvement for the {\tt Cortana} task is comparable to that achieved by doubling the model size. In fact, measured by the amount of training data in our experiments, the model size is not considerably large. We may achieve an even larger gain by increasing the model size further. Finally, we also evaluated the impact of the entropy term in these experiments, and the observation is consistent with that in the previous experiments.  

\begin{table}[t]
  \caption{Experimental results of the larger LSTM model with roughly 63 million parameters.}
  \label{tab:large}
  \centering
  \begin{tabular}{l|ccc}
    \toprule
          & \multicolumn{3}{c}{WER}     \\
     System & {\tt Cortana} & {\tt DMA-O} & {\tt DMA-I} \\
    \hline
    Baseline                 & 9.10        & 17.92 & 12.84                              \\
    {\tt ST} with {\tt H}      & 8.84 & 17.32 & 12.70                      \\
    {\tt ST} without {\tt H} & 8.79 & 17.39 & 12.70\\
    \bottomrule
  \end{tabular}
\end{table}

\section{Future Work}
\label{sec:future}

So far, we have investigated the self-teaching network with frame-level training for speech recognition. In the future, we shall study the case of sequence training, where the posterior probabilities are computed at the sequence-level by running the forward-backward algorithm over lattices. Furthermore, in this work, we only focus on the generalization aspect of the self-teaching network. In the future, we shall also evaluate this approach to train very deep networks, particularly for recurrent networks such as LSTMs. We have shown that the self-teaching network works better for larger models given the fixed amount of training data, which indicates it may achieve larger improvements in the case of lower resource conditions or languages, which will be studied in the future as well. The application of the self-teaching network for other tasks including speaker recognition is also an interesting research direction. 

\section{Conclusions}
\label{sec:conc}

We proposed the self-teaching network to tackle the gradient vanishing problem in deep networks, and to improve the generalization capacity of the network. We approached the problem by introducing the skip connection between the output layer to a lower layer of the network to pass down the supervision signals. We investigated the application of the self-teaching network for speech recognition, where we trained the LSTM based acoustic models using 30 thousand hours of transcribed data. Compared to the closely related ideas such as label smoothing and confidence penalization, we have shown that self-teaching networks can achieve higher and consistent improvements across 4 evaluation datasets, and larger gains were observed with a larger LSTM model.

\bibliographystyle{IEEEtran}

\bibliography{bibtex}

\end{document}